\documentclass[aps,12pt]{revtex4}
\usepackage{epsfig,amsmath,amssymb}

\newcommand{\eq}{{\, \equiv\, }}
\newcommand{\fr}[1]{%   arg1: var name
             \frac{#1}}
\newcommand{\bea}{\begin{eqnarray}}
\newcommand{\eea}{\end{eqnarray}}

\newcommand{\chibar}{\overline{\chi}}

\newcommand{\ket}{{\cal i}}

\newcommand{\bra}{{\cal h}}
\newcommand{\gc}{\bra\fr{\alpha_s}{\pi}G^2\ket}

\newcommand{\ga}{{g_{{\cal A}}}}

\begin{document}
\title{On the Colour Suppressed  \\
 Decay Modes  $\overline{B^0} \rightarrow D_s^+ D_s^-$
and  $\overline{B_s^0} \rightarrow D^+ D^-$}
\author{J. O. Eeg$^a$, S. Fajfer$^{b,c}$, and A. Hiorth$^a$ \\
$^a$Dept. of Physics, Univ. of Oslo, N-0316 Oslo, Norway\\
$^b$Physics Department, University of Ljubljana, Jadranska 19,
SI-1000 Ljubljana, Slovenia\\
$^c$Institut Jo\v{z}ef Stefan, Jamova 39, SI-1000 Ljubljana, Slovenia}
%\author{J. O. Eeg}
\email{j.o.eeg@fys.uio.no}
%\affiliation{Department of Physics, University of Oslo,
%P.O.Box 1048 Blindern, N-0316 Oslo, Norway}
%\author{S. Fajfer}
\email{Svjetlana.Fajfer@ijs.si}
%\affiliation{Istitut Josef Stefan, 
%Jamova 19, SI-1000 Ljubljana, Slovenija}
%\author{A. Hiorth}
\email{aksel.hiorth@fys.uio.no}
%\affiliation{Department of Physics, University of Oslo,
%P.O.Box 1048 Blindern, N-0316 Oslo, Norway}

%\begin{titlepage}
%\pubblock
%\vfill
\vspace{1cm}

\begin{abstract}

We point out that the decay modes
  $\overline{B^0} \rightarrow D_s^+ D_s^-$
and  $\overline{B_s^0} \rightarrow D^+ D^-$
have no factorized contribution.
At quark level these dacays can only proceed through the annihilation
 mechanism, which
in the factorized limit give zero amplitude 
 due to current conservation.
In this paper, we identify the dominating non-factorizable
(colour suppressed)  contributions in terms of 
two  chiral loop contributions and 
one soft gluon emission contribution.
The latter contribution can be 
calculated in terms of the (lowest dimension) gluon condensate within
a recently developed heavy-light chiral quark model.
We find braching ratios $BR (\bar B^0 \to D_s^+ D_s^-)$ 
$ \simeq 7 \times 10^{-5}$ and 
$BR (\bar B^0_s \to D^+ D^-)$ 
$ \simeq 1\times 10^{-3}$.

\end{abstract}

\maketitle

\section{Introduction}
There is presently great interest in decays of $B$-mesons, 
due to numerous experimental results coming from Ba Bar and Belle,
and later at LHC. 

It has been shown \cite{BBNS} that some classes of $B$-meson decay amplitudes
exhibit {\it QCD factorization}.
This means  that, up to $\alpha_s/\pi$ (calculable) and 
$\Lambda_{QCD}/m_b$ (not calculable), 
 their amplitudes factorize into  the product
of two matrix elements of weak currents.
Typically, the decay amplitudes which factorize in this sense
are $B \rightarrow \pi \pi$ and $B \rightarrow K \pi$ where  the energy
 release is big  compared to the light
meson masses.
However, for various decays of the type $\bar{B} \rightarrow D \bar{D}$ where
 the energy release
is of order 1 GeV, QCD factorization is not expected to hold.
(Here $\bar{B}$, $D$, and $\bar{D}$ contain a heavy $b$, $c$, and
anti-$c$
quark respectively).
Such decay modes 
have been considered in
connection with
intermediate $D \bar{D}$ states for other $B$-decay modes \cite{PhaCo}.

In a previous paper \cite{EFZ},
 it was pointed out that the
decay mode $D^0 \rightarrow K^0 \overline{K^0}$ was zero in the factorized 
limit due to current conservation.  However, there are in that case
 non-factorizable (colour suppressed) contributions in terms of chiral
loops and soft gluon emission modelled by a gluon condensate.

\vspace{0.1cm}

In this paper we report on the following observation:
The decay modes
  $\overline{B^0} \rightarrow D_s^+ D_s^-$
and  $\overline{B_s^0} \rightarrow D^+ D^-$
have no factorized (colour non-suppressed) contributions.
At quark level, these decays a priori proceed through the 
annihilation mechanism
$b \bar{s} \rightarrow c \bar{c}$ and $b \bar{d} \rightarrow c \bar{c}$, 
respectively. However, within the factorized limit the annihilation
 mechanism will give a zero amplitude 
 due to current conservation, as for $D^0 \rightarrow K^0 \overline{K^0}$.
But there are non-zero  factorized contributions through the axial
part of the weak  current
 if at least one of $D$-mesons in the final state
is a vector meson $D^*$. Such contributions are, however,
  proportional to the numerically non-favourable Wilson coefficient
  $C_1$, which we will neglect in this short paper.
In contrast, 
 the typical   factorized   decay modes  which proceed
through the spectator mechanism, say
$\overline{B^0} \rightarrow D^+ D_s^-$, 
are proportional to the numerically favourable Wilson coefficient $C_2$. 
If the mesons in this amplitude are also allowed to be vector mesons, 
such amplitudes will 
 generate  non-factorizable ($\sim 1/N_c$) chiral loop contributions
to the process
$\overline{B_d^0} \rightarrow D_s^+ D_s^-$ due to $K^0$-exchange. These will be
  considered in the present paper.

There are also  non-factorizable ($\sim 1/N_c$) contributions
due to soft gluon emission.
Such contributions can be 
calculated in terms of the (lowest dimension) gluon condensate within
a recently developed Heavy Light Chiral Quark Model (HL$\chi$QM) \cite{ahjoe},
which is based on Heavy Quark Effective Theory (HQEFT) \cite{neu}.
This model has been applied to processes with $B$-mesons in \cite{EHP,ahjoeB}.
The gluon condensate contributions is also proportional to
 the favourable Wilson coefficient $C_2$.

In the next section (II), we shortly present the
four quark Lagrangian at quark level. In section III we  present our
 analysis of chiral loop contributions within 
the heavy light chiral perturbation theory. 
In section IV we give the calculation of
non-factorizable
matrix elements due to soft gluons expressed through the (model
dependent) quark condensate. 
 In section V we give the results and 
conclusion.
 Throughout the paper, we will give
formulae and figures for the decay mode 
$\overline{B^0} \rightarrow D_s^+ D_s^-$.
The treatment of  $\overline{B_s^0} \rightarrow D^+  D^-$ will proceed
analogously.

\section{Effective non-leptonic  Lagrangian at quark level}

Based on the electroweak and quantum chromodynamical interactions,
one constructs an effective Lagrangian at
quark level in the standard way: 
 \begin{equation}
 {\cal L}_{W}=  \sum_i  C_i(\mu) \; Q_i (\mu) \; ,
 \label{Lquark}
\end{equation}
where all information of the short distance (SD) loop 
effects above a renormalization scale $\mu$
is contained in the (Wilson) coefficients $C_i$.
 In our case there are two relevant
operators 
\begin{eqnarray}
Q_{1}  = 4  (\overline{q}_L \gamma^\alpha  b_L )  \; 
           ( \overline{c}_L \gamma_\alpha  c_L )  
\qquad  ,  \; \; \; \;
Q_{2}  =  4 \,  ( \overline{c}_L \gamma^\alpha  b_L ) \;  
           ( \overline{q}_L \gamma_\alpha  c_L ) \; ,
\label{Q12} 
\end{eqnarray}  
for $q = d,s$. 
Penguin operators may also contribute, but have small Wilson coefficients.
We may write $C_i = - \frac{G_F}{\sqrt{2}} V_{cb}V_{cq}^* a_i$ ,
where $a_1 \sim 10^{-1}$ and $a_2 \sim 1$ at the scale $\mu = m_b$.
Performing perturbative QCD corrections within
Heavy Quark Effective Theory~(HQEFT)~\cite{neu},
the effective Lagrangian (\ref{Lquark}) can be evolved down to the scale
 $\mu \sim \Lambda_\chi \sim$1 GeV \cite{GKMW,Fleischer}, where one finds 
$|a_1| \simeq 0.4$ and $|a_2| \simeq 1.4$. The $b$, $c$, and
$\overline{c}$
quarks are then treated within HQEFT. 

In order to study non-factorizable contributions at quark level, 
we may use the 
following relation between the generators of $SU(3)_c$ ($i,j,l,n$
are colour indices running from 1 to 3):
\begin{equation}
\delta_{i j}\delta_{l n}  =   \fr{1}{N_c} \delta_{i n} \delta_{l j}
 \; +  \; 2 \; t_{i n}^a \; t_{l j}^a \; ,
\label{fierz}
\end{equation}
where $a$ is the color octet  index.
Then the operators $Q_{1,2}$ may, by means of a Fierz transformation,
be  written in the following way :
\begin{eqnarray}
Q_{1}  =  \frac{1}{N_c} Q_2 + 2 \widetilde{Q_2}
\qquad , \; \; \;    \,Q_{2} =  \frac{1}{N_c} Q_1 + 2 \widetilde{Q_1}
\; \;  , 
\label{QFierz} 
\end{eqnarray}
where 
%the superscript $F$ means ``Fierzed'' and
 the  operators with the ``tilde'' contain colour matrices:
\begin{eqnarray}
\widetilde{Q_{1}}  = 4  (\overline{q}_L \gamma^\alpha t^a  b_L )  \; \,
           ( \overline{c}_L \gamma_\alpha t^a c_L ) \,  
\qquad , \; \;  \;
\widetilde{Q_{2}}  =  4 \,  ( \overline{c}_L \gamma^\alpha t^a b_L )  \; \,
           ( \overline{q}_L \gamma_\alpha t^a c_L ) \; .
\label{QCol} 
\end{eqnarray}  
To obtain
a physical amplitude, one has to calculate the hadronic matrix elements
of the quark operators $Q_i$ within some  framework
describing  long distance (LD) effects. 

\begin{figure}[t]
\begin{center}
   \epsfig{file=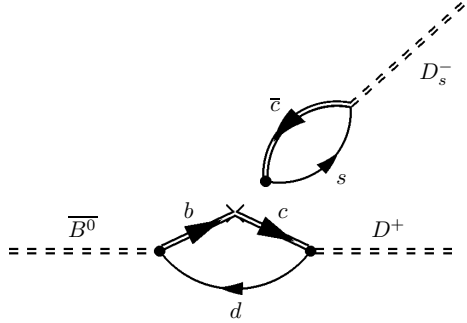,width=7cm}
\caption{Factorized contribution for 
$\overline{B^0}  \rightarrow D^+ D_s^-$
through the spectator mechanism, which does not exist for
 decay mode $\overline{B^0} \rightarrow D_s^+ D_s^-$
we consider in this paper.
 The double dashed lines represent heavy mesons, the double lines
 represent heavy quarks, and the single lines light quarks.}
\label{fig:bdd_fact}
\end{center}
\end{figure}

As an example of a typical {\it factorized} case we choose  
the amplitude for  $\overline{B^0} \rightarrow D^+ D_s^-$ 
obtained from (\ref{Lquark}) and (\ref{Q12}): 
\begin{eqnarray}
\langle D_s^- D^+ | {\cal L}_W| \overline{B^0} \rangle_F \,  = \,
 \; - (C_2 +\frac{1}{N_c} C_1) 
\langle D_s^- | \overline{s}\gamma_\mu \gamma_5 c |0 \rangle
\langle D^+|\overline{c}\gamma_\mu b|\overline{B^0} \rangle \; ,
\label{FactorizedP} 
\end{eqnarray}
which will in section III be compared with our chiral loop contributions.
%The {\it subscript} $F$ means ``factorized''. 
This term is proportional to the $D$-meson decay constant times 
 the Isgur-Wise function
(for $\bar{B} \rightarrow D$ transition) 
and  is vizualized in figure~\ref{fig:bdd_fact}.

\begin{figure}[t]
\begin{center}
   \epsfig{file=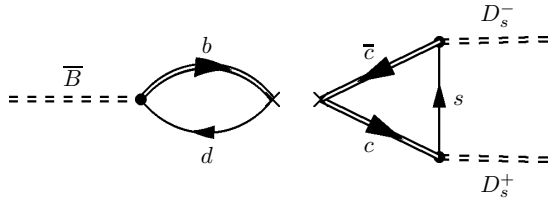,width=8cm}
\caption{Factorized contribution for 
$\overline{B^0}  \rightarrow D_s^+ D_s^-$
through the annihilation  mechanism, which give zero contributions if
both $D_s^+$ and $D_s^-$ are pseudoscalars.}
 \label{fig:bdd_fact2}
\end{center}
\end{figure}

The factorized  amplitude for  $\overline{B^0} \rightarrow D_s^+ D_s^-$ 
obtained from (\ref{Lquark}) and (\ref{Q12}) 
 is vizualized in figure~\ref{fig:bdd_fact2}, and is given by
 \begin{eqnarray}
\langle D_s^- D_s^+ | {\cal L}_W| \overline{B^0} \rangle_F \, = \, 
 4 (C_1 +\frac{1}{N_c} C_2) 
\langle D_s^- D_s^+|\overline{c_L} \gamma_\mu  c_L |0 \rangle
 \langle 0 |\overline{d_L} \gamma^\mu  b_L |\overline{B^0} \rangle \; .
\label{FactorizAnn}
\end{eqnarray}
Unless one or both of the $D$-mesons in the final state 
are vector mesons, this matrix
element is zero due to current conservation:
\bea
\langle D_s^+ D_s^-|\overline{c} \gamma_\mu  c |0 \rangle
 \langle 0 |\overline{d} \gamma^\mu \gamma_5 b |\overline{B^0} \rangle 
 \sim f_B (p_D + p_{\bar{D}})^\mu \,
\langle D_s^+ D_s^- |\overline{c} \gamma_\mu  c |0 \rangle \; = 0 \; .
\eea 

The genuine non-factorizable part for 
 $\overline{B^0} \rightarrow D_s^+ D_s^-$
can  be written 
in terms of  coulored currents  (see eqs. (\ref{QFierz}) and (\ref{QCol})):
\begin{eqnarray}
\langle D_s^- D_s^+ | {\cal L}_W| \overline{B^0} \rangle_{NF} \, =  \,
  8 C_2 \, 
\langle D_s^- D_s^+ | (\overline{q}_L \gamma^\alpha t^a  b_L )  \; 
       ( \overline{c}_L \gamma_\alpha t^a c_L ) \  | \overline{B^0} \rangle
\label{QNFact}
\end{eqnarray}
We observe that the annihilation mechanism amplitude 
in the non-factorizable case has the numerically favourable
 Wilson coefficient $C_2$.
This amplitude is, within the HL$\chi$QM  vizualized later in 
figure~\ref{fig:bdd_nfact2} .

\section{Heavy light chiral perturbation theory}

Our calculations will be based  on 
 HQEFT \cite{neu}, which is a systematic $1/m_Q$ expansion in the
 heavy quark mass $m_Q$.
 The heavy quark field $Q(x) =b(x)$ (eventually $c(x)$ or
 $\overline{c}$)
 is replaced with a ``reduced''
field $Q_v^{(+)}(x)$ for a heavy quark, and $Q_v^{(-)}(x)$ for a heavy
antiquark. These are related to the full
 field $Q(x)$ in the  following way:
\begin{equation}
Q_v^{(\pm)}(x)=P_{\pm}e^{\mp im_Q v \cdot x} Q(x) \, ,
\end{equation}
where $P_\pm$ are projecting operators $P_\pm=(1 \pm \gamma \cdot
 v)/2$.
 The Lagrangian for heavy
quarks is:
\begin{equation}
{\cal L}_{HQEFT} =   \pm \overline{Q_v^{(\pm)}} \, i v \cdot D \, Q_v^{(\pm)} 
 + {\cal O}(m_Q^{- 1}) \; ,
\label{LHQEFT}
\end{equation}
where $D_\mu$ is the covariant derivative containing the gluon field.
In \cite{ahjoeB} the  $1/m_Q$ corrections were calculated for $B
 -\overline{B}$  -mixing. In this paper these will not be considered.

Integrating out the heavy and light quarks, the effective Lagrangian up
to ${\cal O}(m_Q^{-1})$ can be written as 
\cite{itchpt,ahjoe}
\begin{equation}
{\cal L} =  \mp Tr\left[\overline{H^{(\pm)}_{a}}
iv\cdot {\cal D}_{ba}
H^{(\pm)}_{b}\right]\, -\, 
\ga \, Tr\left[\overline{H^{(\pm)}_{a}}H^{(\pm)}_{b}
\gamma_\mu\gamma_5 {\cal A}^\mu_{ba}\right]\, 
,\label{LS1}
\end{equation}
where 
 $H_a^{(\pm)}$ is the heavy meson field  containing
 a spin zero and spin one boson:
\begin{eqnarray}
&H_a^{(\pm)} & \eq  P_{\pm} (P_{a \mu}^{(\pm)} \gamma^\mu -     
i P_{a 5}^{(\pm)} \gamma_5)  \; \; . 
\label{barH}
\end{eqnarray}
The fields $P_M^{(+)}(P_M^{(-)})$ annihilates (creates) a heavy meson
(vector for $M=\mu$ and pseudoscalar for $M=5$) containing a heavy
quark (anti-quark) with velocity  $v$.
Furthermore, $i{\cal D}^\mu_{ba}=i \delta_{ba} D^\mu-{\cal V}^\mu_{ba}$, 
and $a,b$ are flavour indices.
The vector and axial vector fields  ${\cal V}_{\mu}$ and  ${\cal A}_{\mu}$
contain the field 
 $\xi$ (and its first derivative) which is a 3 by 3 matrix containing
 the (would be)  Goldstone octet ($\pi, K, \eta$) :
\begin{equation}
{\cal V}_{\mu}\eq \fr{i}{2}(\xi^\dagger\partial_\mu\xi
+\xi\partial_\mu\xi^\dagger 
) \quad ;  \; \; \;
{\cal A}_\mu\eq  -  \fr{i}{2}
(\xi^\dagger\partial_\mu\xi
-\xi\partial_\mu\xi^\dagger) \quad , \; \; \;  \xi\equiv exp{i(\Pi/f)}\,
\label{defVA}
\end{equation}
where $f$ is the bare pion coupling, and $\Pi$ is a  3 by 3 matrix
which contains the Goldstone bosons $\pi,K,\eta$ in the standard way.
The axial chiral coupling is 
$\ga \simeq 0.6$. Eqs. (\ref{LS1}), (\ref{barH}), and (\ref{defVA}) will be
used for the chiral loop contributions.

The simplest  way to calculate the matrix element of  four quark 
operators like $Q_{1,2}$ in eq. (1) is by inserting vacuum
 states between the two currents, as indicated in section II.
This vacuum insertion approach (VSA)
corresponds to bosonizing the two currents in $Q_{1,2}$ separately
 and multiply them, i.e. the factorized case.
Based on the symmetry of HQEFT,
 the bosonized current for decay of the $b \bar{q}$
system is \cite{itchpt,ahjoe}:
\begin{equation}
 \overline{q_L} \,\gamma^\mu\, Q_{v}^{(+)} \;  \longrightarrow \;
    \fr{\alpha_H}{2} Tr\left[\xi^{\dagger} \gamma^\alpha
L \,  H_{b}^{(+)} \right]
  \; ,
\label{J(0)}
\end{equation}
where $Q_{v}^{(+)}$ is a heavy $b$-quark field, $v$ is its velocity, and
$H_{b}^{(+)}$ is the corresponding heavy meson field.
This bosonization has to be compared with the 
 matrix elements defining  the meson decay 
constants $f_H \, (H=B,D)$ are the same when QCD corrections below
 $m_Q$ are neglected (see \cite{neu,ahjoe}):
\begin{equation}
\alpha_H =  \frac{f_H\sqrt{M_H}}{(C_v + C_\gamma)} \; ,
\label{fb}
\end{equation}
where $C_{v,\gamma}$ are Wilson coefficients due to perturbative QCD 
for scales $\mu < m_Q$
($Q=b,c$ for $H=B,D$). 
We take  $\mu = \Lambda_\chi$, which is the scale  where perturbative
QCD are matched to our hadronic matrix elements.

For the $W$-boson materializing to a $\bar{D}$ we obtain the bosonized current
\begin{equation}
 \overline{q_L} \gamma^\mu\  Q_{\bar{v}}^{(-)} \;  \longrightarrow \;
    \fr{\alpha_H}{2} Tr\left[\xi^\dagger \gamma^\alpha L 
 H_{\bar{c}}^{(-)} \right]
  \; ,
\label{Jqc}
\end{equation}
where $\bar{v}$ is the velocity of the heavy $\bar{c}$ quark and 
$H_{\bar{c}}^{(-)}$ is the corresponding field for the  $\bar{D}$ meson.

For the $b \rightarrow c$ transition, we obtain the bosonized current
\begin{equation}
 \overline{Q_{v}^{(+)}} \,\gamma^\mu\, L Q_{v'}^{(+)}\;  \longrightarrow \;
    - \zeta(\omega) Tr\left[
 \overline{H_c^{(+)}} \gamma^\alpha L  H_{b}^{(+)} \right]
  \; ,
\label{Jbc}
\end{equation}
where $\zeta(\omega)$ is the Isgur-Wise function
 for  the $\bar{B} \rightarrow D$ - transition, and $v'$ is the
 velocity of the heavy $c$-quark.
Furthermore,  $\omega \equiv v \cdot v'= v \cdot \bar{v} = M_B/(2M_D)$.
Note also that 
from conservation of momentum we find 
the relation between the heavy quark velocities: 
\bea
p_{\bar{B}}= p_D + p_{\bar{D}} \; \; \Longrightarrow \; \; 
v^\mu \; = \; \frac{M_D}{M_B} \, (v'+\bar{v})^\mu \; .
\eea

For the weak current for $D \bar{D}$ production  
 (corresponding to the factorizable annihilation mechanism) we obtain
\begin{equation}
 \overline{Q_{v'}^{(+)}} \,\gamma^\mu\, L Q_{\bar{v}}^{(-)}\; 
 \longrightarrow \;
   - \zeta(-\lambda) Tr\left[
 \overline{H_c^{(+)}} \gamma^\alpha L  H_{\bar{c}}^{(-)} \right]
  \; ,
\label{Jcc}
\end{equation}
where $\lambda= \bar{v} \cdot v' = [M_B^2/(2M_D^2) -1]$.
The Isgur-Wise function $\zeta(-\lambda)$ in (\ref{Jcc}) is a complex 
function, and not so well-known as for the $b \rightarrow c$
transition.
 In the factorized limit, the matrix elements of the four quark
operators are obtained by multiplying the bosonized currents above.

In the following we will consider explicitely the decay mode
$\overline{B^0} \rightarrow D_s^+  D_s^-$. The analysis of
$\overline{B^0_s} \rightarrow D^+  D^-$ proceed the same way.
To calculate the chiral loop amplitudes
 we need the (factorized) amplitudes for 
$\overline{B_s^{*0}} \rightarrow D_s^+ D^{*-}$ and
$\overline{B^0} \rightarrow D^{*+} D^{*-}$, which proceed through the
spectator mechanism as in  figure~\ref{fig:bdd_fact}.
The point is that the leading chiral coupling obtained from (\ref{LS1})
is a coupling between a pseudoscalar meson $H$, vector meson $H^*$ 
a light pseudoscalar $M$ ($=\pi, K, \eta$).
\begin{figure}[t]
\begin{center}
   \epsfig{file=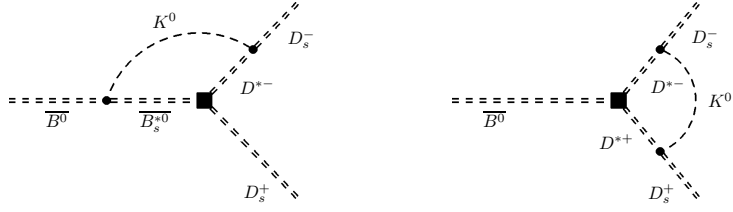,width=10cm}
\caption{Non-factorizable chiral loops for 
$\overline{B^0} \rightarrow D_s^+ D_s^-$.}
\label{fig:chiral1}
\end{center}
\end{figure}
Using the bosonized currents in eqs. (\ref{Jqc}) and (\ref{Jbc}),
we obtain the following chiral loop amplitude for the process
 $\overline{B^0} \rightarrow D_s^+ D_s^-$ 
from the figure~\ref{fig:chiral1}:
\bea
A(\overline{B^0} \rightarrow D_s^+ D_s^-)_\chi \, = \,  
\left(V_{cd}^*/V_{cs}^*\right)
 \, A(\overline{B_d^0} \rightarrow D_d^+ D_s^-)_F \cdot
\; R^\chi \; \, ,
\label{chiralR}
\eea
where the  factorized amplitude
for the process $\overline{B^0} \rightarrow D^+ D_s^-$ is
\bea
A(\overline{B^0} \rightarrow D^+ D_s^-)_F = 
 -\frac{G_F}{{\sqrt 2}}V_{cb} V_{cs}^*
 \, a_2 \,\zeta(\omega) f_D M_D 
\sqrt{M_B M_D} \; (\lambda + \omega) \; \, .
\label{BDDF}
\eea
The quantity $R^\chi$ is a sum of contributions $R^\chi_{1,2}$ 
from the left and right
part of figure~\ref{fig:chiral1} respectively.
In the $\overline{MS}$ scheme the results  for $R^\chi_{1,2}$
 are
\begin{eqnarray}
R^\chi_1 \; = \; 
&&\fr{m_K^2}{2 (4\pi f)^2} \ga^2\left[\left\{2 
\frac{(\omega+1)}{(\omega+\lambda)}\,r(-\omega)- 1\right\}
\ln\left(\fr{m_K^2}{\Lambda_\chi^2}\right)- 1\right] \;  , \\
R^\chi_2 \; = \; 
&&\fr{m_K^2}{2 (4\pi f)^2} \ga^2 \left[\left\{2 
\frac{(\omega+1)}{(\lambda+\omega)}\,r(-\lambda)- 1\right\}
\ln\left(\fr{m_K^2}{\Lambda_\chi^2}\right)- 1\right] \; .
\end{eqnarray}
Adding these two contributions we find :
\begin{eqnarray}
R^\chi \; = \;
&&\fr{m_K^2}{(4\pi f)^2}\ga^2\left[\left\{
\frac{(\omega+1)}{(\omega+\lambda)}\, [r(-\omega)+r(-\lambda)]
- 1\right\}
\ln\left(\fr{m_K^2}{\Lambda_\chi^2}\right)- 1 \right]
\label{chiralT}
\end{eqnarray}
As usual, the $1/N_c$ suppression is due to $f^2 \sim N_c$.
The function $r(x)$ is also appearing in loop calculations
\cite{GKMW,Fleischer} of the
anomalous dimension in HQEFT  (for $x > 1$ and $x< -1$ respectively):
\begin{equation}
r(x) \equiv \fr{1}{\sqrt{x^2-1}}\, 
\text{ln}\left(x+\sqrt{x^2-1}\right) \quad , \; \; \; 
r(-x)= -r(x) + \fr{i \pi}{\sqrt{x^2-1}} \, \; ,
\end{equation}
which  means that the amplitude
 gets an imaginary part.
Numerically, we find
\bea
R^\chi \;\simeq \; 0.12 - 0.26 i  \; .
\label{XNum}
\eea

\section{Non-factorizable soft gluon emission }

\begin{figure}[t]
\begin{center}
   \epsfig{file=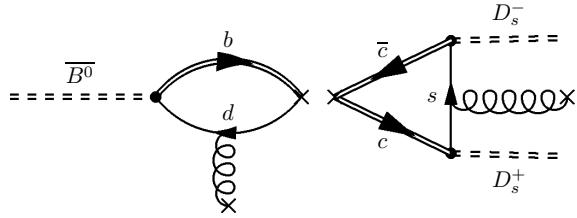,width=8cm}
\caption{Non-factorizable contribution for 
$\overline{B^0}  \rightarrow D_s^+ D_s^-$
through the annihilation mechanism with additional soft gluon emision.
 The wavy lines represent soft
gluons ending in vacuum to make gluon condensates.}
\label{fig:bdd_nfact2}
\end{center}
\end{figure}

The genuine non-factorizable part  (see eqs. (\ref{QFierz}),
(\ref{QCol}) and (\ref{QNFact}) )
 can,  within the framework presented in this section, be written in a
 quasi-factorized way
in terms of matrix elements of coulored currents:
\begin{eqnarray}
\langle D_s^+ D_s^- | {\cal L}_W| \overline{B^0} \rangle_{NF}^G \, = \,
  8 C_2 \, 
\langle D_s^+ D_s^-|\overline{c}_L \gamma_\mu t^a c_L |G \rangle
 \langle G |\overline{d}_L \gamma^\mu  t^a b_L |\overline{B^0} \rangle
\; ,
\label{QGlue}
\end{eqnarray}
where a $G$ in the bra-kets symbolizes emision of one gluon 
(from each current) as vizualized in figure  \ref{fig:bdd_nfact2}.  
We observe that the annihilation mechanism amplitude 
in the 
non-factorizable case has the numerically favourable Wilson coefficient $C_2$.

In order to calculate the matrix elements (\ref{QGlue}), we will use 
a model which incorporates emision of soft gluons modelled by a gluon
condensate. This will be the Heavy Light
Chiral Quark Model (HL$\chi$QM) recently developed in \cite{ahjoe}.
This model belongs to a class of models extensively studied in the
literature \cite{chiqm,barhi,itCQM,effr,ebert3,bijnes}.
 For details we refer to ref. \cite{ahjoe}.

The Lagrangian for the HL$\chi$QM is 
\begin{equation}
{\cal L}_{\text{HL$\chi$QM}} =  {\cal L}_{\text{HQEFT}} +  {\cal L}_{\chi\text{QM}}  +   {\cal L}_{\text{Int}} \; .
\label{totlag}
\end{equation}
The first term is given in equation (\ref{LHQEFT}).
The light quark sector is described by the Chiral Quark Model ($\chi$QM),
having a standard QCD term and a term describing interactions between
quarks and  (Goldstone) mesons: 
\begin{equation}
{\cal L}_{\chi\text{QM}} =  
\chibar \left[\gamma^\mu (i D_\mu   +    {\cal V}_{\mu}  +  
\gamma_5  {\cal A}_{\mu})    -    m \right]\chi  \; .
\label{chqmR}
\end{equation}
Here $m$ is the SU(3) invariant constituent light quark mass, and
 $\chi$ is the flavour rotated quark fields given by
$\chi_L  =   \xi^\dagger q_L \; \; , \;  \chi_R  =   \xi q_R$, 
where $q^T  =  (u,d,s)$ are the light quark fields. The left- and
 right-handed
 projections $q_L$ and $q_R$ are transforming after $SU(3)_L$ and $SU(3)_R$
respectively. In (\ref{chqmR}) we have discarded terms involving the light
 current quark mass which is irrelevant in the present paper.
The covariant derivative $D_\mu$ in (\ref{chqmR})
contains the soft gluon field forming the gluon condensates. The gluon 
condensate contributions are calculated by Feynman diagram techniques
as in \cite{ahjoe,EHP,ahjoeB,epb,BEF}. They may also be calculated
by means of heat kernel techniques as in \cite{pider,bijnes,ebert3}.

The interaction between heavy meson fields and heavy quarks are
described by the following Lagrangian \cite{ahjoe}:
\begin{equation}
{\cal L}_{Int}  =   
 -   G_H \, \left[ \chibar_a \, \overline{H_a^{(\pm)}} 
\, Q^{(\pm)}_{v} \,
  +     \overline{Q_{v}^{(\pm)}} \, H_a^{(\pm)} \, \chi_a \right] \; ,
\label{Int}
\end{equation}
where
 $G_H \sim \sqrt{2 m}/f$ is a  coupling constant.
In \cite{ahjoe} it was shown
how (\ref{LS1}) could be obtained from the HL$\chi$QM.
Performing this bosonization of the HL$\chi$QM,  one encounters divergent loop
integrals which will in general be quadratic-, linear- and
logarithmic divergent \cite{ahjoe}.
 Also, as in the
light sector \cite{BEF} the quadratic and logarithmic integrals are
related to the quark condensate and the gluon condensate respectively.

Within the model, one finds the following expression for the Isgur-Wise
function \cite{ahjoe}
\begin{equation}\label{iw}
\zeta(\omega)= \fr{2}{1+\omega} \left(1-\rho \right)+
 \rho \, r(\omega)\, , 
\end{equation}
where $\rho$ is a hadronic parameter giving the deviation from the
leading value \cite{ahjoe}:
\begin{equation}
G_H^2  =   \fr{2m}{f^2} \, \rho \; \, ,
\quad 
\rho \, \eq \, \fr{(1 +  3g_{\cal A}) + 
 \frac{\mu_G^2}{\eta \, m^2}}{4 (1 + \frac{N_c m^2}{8 \pi f^2} )}
\quad , \; \; \; \mu_G^2(H) \, =  \, \frac{3}{2} m_Q (M_{H^*} -  M_H) ,
\label{rho}
\end{equation}
where $\eta= (1+ 2/\pi)$. Numerically, the deviation of $\rho$
 one is of order $10\%$. 
The simple expression in (\ref{iw}) is
modified by perturbative QCD corrections down to 
$\mu= \Lambda_\chi$ (analogous to eq. (\ref{fb})) and chiral loop corrections. 
Our model dependent expression $\zeta(\omega)$ in (\ref{iw}) give a
 good description of the Isgur-Wise function.

 The  left part in figure~\ref{fig:bdd_fact2} with gluon emision
 gives us the bosonized coloured current :
\bea
 \left(\overline{q_L} \, t^a  \,\gamma^\alpha \, Q_{v_B}^{(+)}\right)_{1G} 
\;   \longrightarrow \;  
- \fr{G_H \, g_s}{64 \pi} \,G_{\mu\nu}^a
\;Tr\left[\xi^\dagger \gamma^\alpha  L \, H_b^{(+)}
\left( \sigma^{\mu\nu} \, - \,
 \fr{2 \pi f^2}{m^2 \, N_c}  \{\sigma^{\mu\nu},
 \gamma \cdot v \} \,
 \right)\right] \; ,
\label{1G}
\eea
where $G^a_{\mu \nu}$ is the octet gluon tensor, and
 $H_b^{(+)}$ represents the heavy $\bar{B}$-meson fields.
Similarly the (heavy) $D$- and $\bar{D}$-mesons are represented by
$H_c^{(+)}$ and $H_{\bar{c}}^{(-)}$ corresponding to a heavy quark field
$Q_{v'}^{(+)}$ and heavy anti-quak field $Q_{\bar{v}}^{(-)}$ respectively.
The symbol $\{\; , \; \}$  denotes the anti-commutator.

For the creation of a $D \bar{D}$ pair in the right part of figure 
\ref{fig:bdd_fact2}, the analogue of (\ref{1G}) is
\bea
 \left(\overline{Q_{v'}^{(+)}} \,t^a \; 
\gamma^\alpha \, L Q_{\bar{v}}^{(-)}\right)_{1G} \;
\;   \longrightarrow \;  
 \fr{G_H^2 \, g_s}{32 \pi}\,G_{\mu\nu}^a
\;Tr\left[ \overline{H_c^{(+)}} \; \gamma^\alpha \, L \,  
  H_{\bar{c}}^{(-)}  \right.  \nonumber \\
\left.  \times  \left( \frac{\widetilde{r}}{\pi} \sigma^{\mu\nu}
\, + \, \frac{1}{4 m (\lambda-1)}
 \left\{\sigma^{\mu\nu}, \gamma \cdot t \right\}  \,
  \right)\right] \; ,
\label{1Gc}
\eea
where $\, t=v'-\bar{v} \,$, and $\widetilde{r} \equiv r(-\lambda)$.
Multiplying the currents in eqs. (\ref{1G}) and (\ref{1Gc}),
and using   the replacement:
\begin{equation}
g_s^2 G_{\mu \nu}^a G_{\alpha \beta}^a  \; \rightarrow 4 \pi^2
 \gc \frac{1}{12} (g_{\mu \alpha} g_{\nu \beta} -  
g_{\mu \beta} g_{\nu \alpha} ) \, ,
\label{gluecond}
\end{equation}
we obtain the bosonized version for the operator $\widetilde{Q_1}$
in eq. (\ref{QCol}) (see also eq. (\ref{QGlue})) as the product of two
traces. (The expression may be simplified by using the Dirac algebra, but we do
not enter these details here). 

Taking the pseudoscalar parts of (\ref{1G}) and (\ref{1Gc}),
we find the gluon condensate contribution
for   $\overline{B^0} \rightarrow D_s^+ D_s^-$
within our model:
\bea
A(\overline{B^0} \rightarrow D_s^+ D_s^-)_G =  
  C_2  \gc \; 
\fr{(G_H \sqrt{M_B})^3}{384 m} 
 \left(1 + \,
\frac{3 \, \widetilde{r}}{\pi} \right) \; .
\label{B2DD}
\eea
(For our algebraic manipulations, the
  program FORM \cite{Verm} was useful).
The ratio between this  amplitude and the factorized one in
(\ref{BDDF}) scales as $M_D/(N_c M_B)$ times
hadronic parameters calculated within HL$\chi$QM. 
We define a quantity $R^G$ for the gluon
condensate amplitude  analogously to  $R^\chi$ in (\ref{chiralR})
 and (\ref{chiralT})
for chiral loops.
Numerically, we find that the ratio between the two amplitudes
 in (\ref{B2DD}) and (\ref{BDDF}) is
\bea
R_G \; \simeq \; 0.055  + 0.16 i  \; ,
\label{RNum}
\eea
which is of order one third of the chiral loop contribution in
eq. (\ref{chiralT}).

\section{ Discussion and Results}

Our amplitude is complex as expected. In the chiral loop amplitude
these are due to physical cuts (exchanges of physical particles)
to the one-loop order we consider in this paper.
The Wilson coefficients turn complex when the
c-quark is treated  \cite{GKMW,Fleischer} within HQEFT. This is also the
case for the matrix elements that
these Wilson coefficients  should be matched to.
There is  a potential problem with a quark model without
confinement that the amplitude
may get an imaginary part due to production of free quarks.
Still, within HQEFT one can hardly
 distinguish $m_c$ from $M_D$ because of the reparametrization
invariance.
Thus, at the present stage, it is not clear
how well our model describes imaginary matrix elements, and we will
not go into such details here, as the numerical consequences turn out
to be minor.

Adding the amplitudes $R_\chi$ and $R_G$ and multiplying with the
Wilson coefficient \cite{GKMW,Fleischer} $a_2 \simeq 1.33 + 0.2 i$, 
we obtain the quantity:
\bea
\label{Tres}
\widetilde{R_T} \, \equiv \, 
a_2 \, (R_\chi + R_G) \; \simeq 0.26 - 0.11 i \; .
\eea
Dropping  the imaginary parts of the three quantities would give
instead the value $\simeq 0.25$.  Anyway, 
we have found that the amplitude for
$\overline{B^0} \rightarrow D_s^+ D_s^-$ is of order $15-20 \%$
of the factorizable amplitude for
$\overline{B^0} \rightarrow D^+ D_s^-$, before the different KM-factors are
taken into account.
We  obtain the branching ratios
\bea
\label{Brd}
 BR(\overline{B^0_d} \to D_s^+ D_s^-) \,  = \,  
 6.5 \times 10^{-5} \times |\frac{V_{cb}}{0.041} \, \frac{V^*_{cd}}{0.223}
\, \frac{\widetilde{R_T}}{0.25}\, \frac{\zeta(\omega)}{0.9} |^2
\eea
and 
\begin{equation}
\label{Brs}
 BR(\overline{B^0_s} \to D^+ D^-) \, = \, 8.9 \times 10^{-4} \times  
 |\frac{V_{cb}}{0.041} \, \frac{V^*_{cs}}{0.974} \, 
\frac{\widetilde{R_T}}{0.25} \,  \frac{\zeta(\omega)}{0.9} |^2
\end{equation}
The difference between the branching ratios is mainly due to the difference in
KM factor. 
Taking into account the comments above, we end up with the conclusion that
\bea
\label{BR}
 BR(\overline{B^0_d} \to D_s^+ D_s^-) \simeq 
 7 \times 10^{-5}  \; \; , \text{and} \quad
 BR(\overline{B^0_s} \to D_s^+ D_s^-) \simeq 1 \times 10^{-3} \; \; .
\eea

The ongoing searches at Belle might soon give  the limit on the rate 
 $\overline{B^0} \to D_s^+ D_s^-$, while the detection of the 
$\overline{B_s^0}$ mode might  be  presently 
 more difficult due to troubles with $\overline{B_s^0}$  identification.

\vspace{1cm}
S.F. thanks P. Kri\v{z}an and B. Golob 
for fruitful discussion

\bibliographystyle{unsrt}

\begin{thebibliography}{99}

%\bibitem{Exp}
%Experimental reference?

\bibitem{BBNS}
M. Beneke, G. Buchalla, M. Neubert, C.T. Sachrajda {\it Phys. Rev.
Lett.} {\bf 83} (1999) 1914.
%-1917. 

\bibitem{PhaCo} 
P. Colangelo, F. De Fazio, and T.N. Pham,  
{\it  Phys. Lett.} {\bf B 542} (2002) 71. \\
%-79.\\
C. Isola, M. Ladisa, G. Nardulli, T.N. Pham, and P. Santorelli,\\  
{\it  Phys. Rev.} {\bf D 65} (2002) 094005.


\bibitem{EFZ}
J.O. Eeg, S. Fajfer, J. Zupan, {\it Phys. Rev. } {\bf D 64} (2001)  034010.  

\bibitem{ahjoe} A.~Hiorth and J.~O.~Eeg, {\it Phys. Rev.} {\bf D 66} 
(2002) 074001.


\bibitem{neu} For a review, see
 M.~Neubert, {\it Phys. Rep.} {\bf 245} (1994) 259.

\bibitem{EHP}
 J.~O.~Eeg, A.~Hiorth, A.~D.~Polosa, 
{\it Phys. Rev.} {\bf D 65} (2002) 054030. 

\bibitem{ahjoeB}
 A.~Hiorth and J.~O.~Eeg, 
{\it Eur.Phys.J.direct} {\bf C30} (2003) 006.

\bibitem{GKMW}
B. Grinstein, W.~Kilian, T.~Mannel, and M.B. Wise, {\it Nucl. Phys.} {\bf B363}
(1991) 19.
%-33.

\bibitem{Fleischer}
R. Fleischer, {\it Nucl. Phys.} {\bf B 412} (1994) 201.
%-224.


\bibitem{itchpt} R.~Casalbuoni, A.~Deandrea, N.~Di~Bartelomeo, 
R.~Gatto, F.~Feruglio and  G.~Nardulli, \\
 {\it Phys. Rep.} {\bf 281} (1997) 145.\\
J.L. Goity,, {\it Phys. Rev.} {\bf D46} (1992) 3929.
%-3936 .


\bibitem{chiqm}
J.~A.~Cronin, {\it Phys. Rev.} {\bf 161} (1967) 1483, \\
S.~Weinberg, {\it Physica} {\bf  96A} (1979) 327, \\
D.~Ebert and M.K.~Volkov, {\it Z. Phys.}  {\bf C 16}(1983) 205, \\
A.~Manohar and H.~Georgi, {\it Nucl. Phys.}  {\bf B234}(1984) 189, \\
J.~Bijnens, H.~Sonoda and M.B.~Wise, {\it Can. J. Phys.} {\bf 64} (1986) 1, \\
D.~Ebert and  H.~Reinhardt, {\it Nucl. Phys.}  {\bf B71}(1986) 188, \\
D.~I.~Diakonov, V.~Yu.~Petrov and P.~V.~Pobylitsa,
 {\it Nucl. Phys.} {\bf B306}(1988) 809, \\
D.~Espriu, E.~de~Rafael and J.~Taron, {\it Nucl. Phys.} {\bf B345} (1990) 22.

\bibitem{barhi}
W.~A.~Bardeen and C.~T.~Hill, {\it Phys. Rev.} {\bf D49} (1994) 409 .

\bibitem{itCQM}
A.~Deandrea, N.~Di~Bartelomeo, R.~Gatto, G.~Nardulli, and A.~D.~Polosa, \\
{\it Phys. Rev.} {\bf D58} (1998) 034004. 
  A.~Polosa, {\it  Riv. Nuovo Ceimento} {\bf 23 N11} (2000) 1.  

\bibitem{effr}
D.~Ebert, T.~Feldmann  R.~Friedrich and H.~Reinhardt,
{\it Nucl. Phys.} {\bf B434} (1995) 619.


\bibitem{ebert3} D.~Ebert and M.K.~Volkov, {\it Phys.Lett.} {\bf B 272}
(1991), 86.


\bibitem{bijnes} J.~Bijnens, {\it Phys. Rept.} {\bf 265} (1996).


\bibitem{epb}
J.~O.~Eeg and I.~Picek, {\it Phys. Lett.} {\bf B301} (1993) 423, 
%J.~O.~Eeg and I.~Picek, {\it Phys. Lett.}
{\it ibid.} {\bf B323} (1994) 193, \\
A.E.~Bergan and J.O.~Eeg, {\it Phys. Lett.} {\bf390} (1997) 420.
 

\bibitem{BEF}
S.~Bertolini, J.O.~Eeg and M.~Fabbrichesi, 
{\it Nucl. Phys.} {\bf B449} (1995) 197, \\
V.~Antonelli, S.~Bertolini, J.O.~Eeg,
M.~Fabbrichesi and E.I.~Lashin, \\
 {\it Nucl. Phys.} {\bf B469} (1996) 143, 
%\\ 
S.~Bertolini, J.O.~Eeg,
M.~Fabbrichesi and E.I.~Lashin, \\
 {\it Nucl. Phys.} {\bf B514} (1998) 63,
% \\
{\it ibid} {\bf B514} (1998) 93.


\bibitem{pider}
A.~Pich and E.~de Rafael, Nucl. Phys. {\bf B358} (1991) 311.


\bibitem{Verm}
J.A.M. Vermaseren, ``Symbloic Manipulation with  FORM'', \\
CAN 1991, Amsterdam. (ISBN 90-74116-01-9). 


\end{thebibliography}

\end{document}